\documentclass[nonacm]{acmart}

\usepackage[utf8]{inputenc}
\usepackage{url}
\usepackage{graphicx}
\usepackage{enumitem}
\usepackage[framemethod=TikZ]{mdframed}

\newenvironment{defframe}[0]
  {\mdfsetup{
    frametitleaboveskip=-\ht\strutbox,
    frametitlealignment=\center,
    backgroundcolor=gray!10,
    middlelinewidth=1pt,
    roundcorner=3pt
    }
  \begin{mdframed}[nobreak=true]%
  \centering
  }
  {\end{mdframed}}

\title{Introduction to Digital Twins for the Smart Grid}
\subtitle{Book Chapter in ``Digital twin technology and smart grid'' (Elsevier, 2026)}

\author{Xiaoran Liu}
\orcid{0009-0000-9908-7406}
\affiliation{
  \institution{McMaster University}
  \city{Hamilton}
  \country{Canada}
}
\email{liu2706@mcmaster.ca}

\author{Istvan David}
\orcid{0000-0002-4870-8433}
\affiliation{
  \institution{McMaster University}
  \city{Hamilton}
  \country{Canada}
}
\affiliation{
  \institution{McMaster’s Centre for Software Certification (McSCert)}
  \city{Hamilton}
  \country{Canada}
}
\email{istvan.david@mcmaster.ca}

\renewcommand{\shortauthors}{Liu and David}

\begin{document}

\maketitle

\section{Foundations and definitions}\label{sec:foundations}

\textit{Digital twins} are real-time and high-fidelity virtual representations of physical assets~\cite{rasheed2020digital}, referred to as the \textit{physical twin}~\cite{kritzinger2018digital} or \textit{actual twin}~\cite{tao2019digital}. Digital twins reflect the prevalent state of the physical twin and offer cost-efficient and safe alternatives for interacting with it.
To avoid confusion and whenever necessary, collections of digital and physical twins, collectively, are referred to as \textit{digital twin systems}.

A unique trait of a digital twin system is the strong bi-directional coupling between the digital and physical twin in which the digital twin controls the physical twin through computational reflection---the behavior exhibited by a reflective system based on a causal connection to the reflected system~\cite{maes1987concepts}. In digital twin systems, such a causal connection is typically achieved through the digital twin continuously processing sensor data from the physical twin. Computational reflection of the physical twin is what allows the digital twin to reason about the current and future states of the physical twin and control it in accordance with optimality, safety, and other goals of interest.

\begin{figure}[h]
    \centering
    \includegraphics[width=0.54\linewidth]{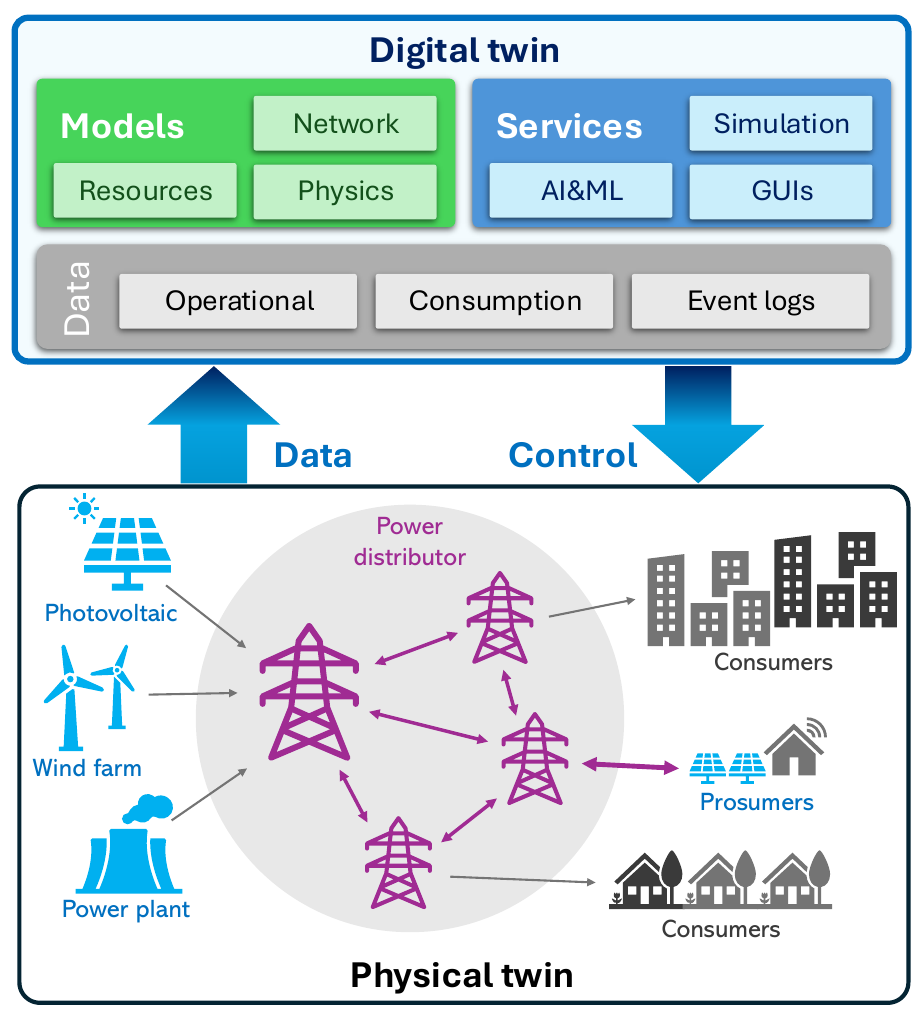}
    \caption{A typical digital twin of a power grid}
    \label{fig:dt-powergrid}
\end{figure}

These concepts are shown in Figure~\ref{fig:dt-powergrid}. The digital twin contains various models of the grid, capturing, for example, network dynamics, resource utilization, and environmental conditions (e.g., temperature, humidity, light conditions). These models are maintained through real-time data acquisition. Services make use of models, e.g., to simulate future states of the grid, and to present the results of such simulations to end-users through graphical user interfaces.
Models in the digital twin can be inclusive of the power distribution network, the consumer and prosumer (power consumers who also produce and contribute power) points, as well as the primary power generators, e.g., power plants, wind farms, and photovoltaic devices. The more subsystems the digital twin covers, the more control it may have over the grid.

\subsection*{Brief history of digital twins}

The concept of the digital twin is rooted in NASA's Apollo programs, tracing back to the 1960s. In response to spacecraft failures, NASA envisioned ``living models'' to support its Apollo missions~\cite{allen2021digital}, i.e., simulators that evaluate the health of the spacecraft based on continuously ingested data. One could consider this one of the first digital twins. A few decades later, in the early 2000s, driven by advanced product lifecycle management (PLM) ambitions, Grieves~\cite{grieves2017digital}
suggested a similar mechanism to develop the ``conceptual ideal for PLM'', later labeled as ``the information mirroring model.''
The notion of a digital twin has been eventually defined in a 2010 NASA roadmap document~\cite{shafto2010draft} as follows: ``an integrated multiphysics, multiscale,
probabilistic simulation of an as-built vehicle or system that uses the best available physical models, sensor updates, fleet history, etc., to mirror the life of its corresponding flying twin.'' The digital twin is also considered ``ultra-realistic,'' thanks to high-fidelity physical models, real-time sensor data, and the ability to aggregate maintenance history and fleet data.
In their hype cycle report for emerging technologies, Gartner reported in 2017 that digital twins entered the innovation trigger phase, i.e., a rapid growth of interest in the concept and subsequent adoption is imminent in the next 5--10 years.\footnote{\url{https://www.gartner.com/en/documents/3768572}} Due to an unusually rapid progress, the 2018 edition of the same report already placed digital twins at the peak level of (inflated) expectations.\footnote{\url{https://www.gartner.com/en/documents/3885468}} Later reports do no longer mention digital twins, and it is widely assumed that digital twins have reached the steady level of maturity, i.e., the so-called plateau of productivity~\cite{paredis2022towards}.
Indeed, today, digital twins are drivers to complex and mature solutions. Some of the key applications of digital twins include automated optimization, real-time reconfiguration, and intelligent adaptation of physical systems. These capabilities and the simplicity of the digital twin concept render digital twins particularly appealing in settings where cyber-physical systems have to be controlled in a computer-automated fashion, such as manufacturing~\cite{touhid2023building}, uncrewed avionics~\cite{lei2021intelligent}, smart ecosystems~\cite{michael2024digital}, and smart grids~\cite{jafari2023review}.

\subsection*{Why digital twins matter?}\label{sec:why-dts-matter}

Digital twins provide a substantial competitive advantage across industries. Capgemini reports an average of 13\% decrease in costs and a 15\% increase in operational efficiency across organizations implementing digital twins.\footnote{\url{https://www.capgemini.com/be-en/insights/research-library/digital-twins/}} McKinsey reports these numbers being in the 20\%--25\% range in warehouse operations\footnote{\url{https://www.mckinsey.com/capabilities/operations/our-insights/improving-warehouse-operations-digitally}} with additional 30\%--50\% reductions in machine downtime.\footnote{\url{https://www.mckinsey.com/capabilities/operations/our-insights/capturing-the-true-value-of-industry-four-point-zero}}
Thus, the business case for digital twins is rather clear. In many domains, digital twins are key drivers of digital transformation---the process of organizations adapting to changes in their business through utilizing digital technology~\cite{vial2019understanding}. As companies progress through their digital transformation journeys, they adjust their value-creation processes and strategies, largely driven by the opportunities in digitalization~\cite{kraus2021digital}.
Not only do digital twins enable digital transformation in many domains, but the endeavor itself through which companies develop and deploy digital twins often also enforces the digital transformation of the company.
This is because advanced digitalization requires the acquisition of new skills, tools, licenses, and best practices. Even domains that have been least amenable to digitalization---such as agronomy and healthcare---adopt digital twin technology at a rapid pace~\cite{david2023automated}.

Another foundational enabler of digital transformation, the digital thread~\cite{margaria2019digital} makes extensive use of digital twin technology, too. The digital thread is a record of a product's or system's lifetime from its creation to its retirement. Digital twins act as real-time data collection facilities across the entire lifecycle and as contextualized representations that continuously reflect the current state and behavior of physical assets. This real-time mirroring allows the digital thread to stay up-to-date and allows for informed decision-making throughout the value chain.
For example, in a smart grid, the digital thread might capture the operational history of a wind turbine, ranging from its design parameters, through its manufacturing and installation records, to real-time performance data and maintenance logs. When the digital twin detects physical indications of wear and tear, this information is recorded in the digital thread, allowing for predictive maintenance and repair. This rich information contained in the digital thread allows quicker repair and maintenance operations as technicians can access and correlate error diagnostics with specifications.

In addition to driving digital transformation, digital twins excel in driving sustainability transition journeys, too. Today, we see an increasing interest in aligning digital transformation principles with corporate sustainability ambitions~\cite{lazazzara2024digital}. Industry 4.0 is being gradually replaced by Industry 5.0, an improvement that puts research and innovation at the service of the transition to a sustainable, human-centric and resilient industry.\footnote{\url{https://research-and-innovation.ec.europa.eu/research-area/industrial-research-and-innovation/industry-50\_en}} By integrating AI, automation, and robotic technology into manufacturing production, while enabling workers to play an active role in decision making (i.e., promoting on human creativity), Industry 5.0 concentrates on value-added tasks, while automating repetitive and monotonous tasks. Digital twins serve as the platform for the integration of various technological components and allow humans to interface with them in real-time. By that, digital twins also foster circularity in systems engineering~\cite{david2024circular}.

\subsection*{Definitions: what \textit{is} and what \textit{is not} a digital twin?}

\begin{figure}[h]
    \centering
    \includegraphics[width=0.5\textwidth]{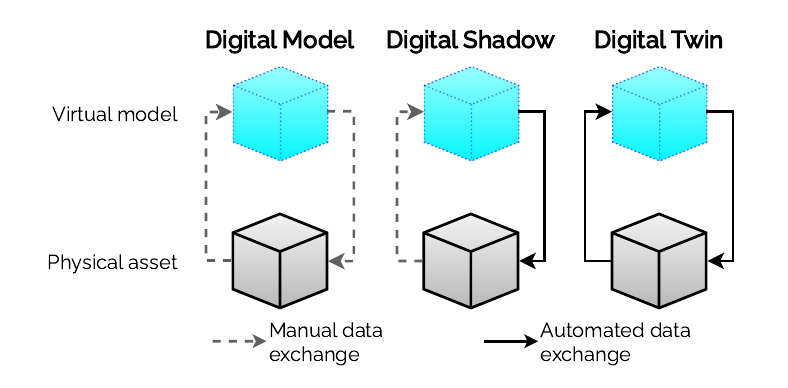}
    \caption{Digital assets classified by the level of automation in data exchange with the physical asset, as per \citet{kritzinger2018digital}}
    \label{fig:dt-types}
\end{figure}

A common typology of digital twin systems is due to \citet{kritzinger2018digital}, who distinguish between three types of systems, depending on the level of automation in data exchange and control (Figure~\ref{fig:dt-types}). \textit{Digital twins} process data from the physical twin in an autonomous fashion, and possess the ability to control them. A system is called a \textit{digital shadow} when automated control cannot be guaranteed, but still processes data from the physical system continuously. Finally, a \textit{digital model} is updated with data manually while having no control over the physical system.

Quite often, digital shadows are generously labeled as digital twins, even though this is not entirely correct.
In a somewhat more detailed view, \citet{tao2019digital} suggest \textit{services} and \textit{data} to be promoted to first-class components in digital twin systems. In their model, data takes the central role through with physical and virtual entities are connected, and based on which services are provided by the digital twin.
In reaction to the often overlooked human aspects of digital twins, \citet{david2024infonomics} suggest the separation of the ability and liberty of a digital twin to act autonomously. In this classification, \textit{human-supervised digital twin}s are able to control physical systems but they are not allowed to. To shift towards fully automated digital twins, human factors, especially trust in the digital twin needs to improve. In contrast, \textit{human-actuated digital twin}s are unable to control physical twins in an automated fashion even though this capability would be needed and instead, they rely on human actuation, e.g., by issuing work commands to humans.

More industry-attuned definitions are provided by companies and suppliers that provide digital twin platforms or use them for innovative services.
Some of these are well-aligned with the usual definitions of digital twins.
For example, NVIDIA defines digital twins as ``\textit{virtual representations of products, processes, and facilities that enterprises use to design, simulate, and operate their physical counterparts}.''\footnote{\url{https://www.nvidia.com/en-eu/glossary/digital-twin/}} An interesting element in this definition are processes which, apparently, can be twinned, too. This is not as surprising as it seems at first. When twinning a process, a digital twin reflects a real-life process based on its observable and measurable outputs; and controls the process based on simulations about its future states.
Other industry actors focus more on the extensions of the digital twin concept through their own unique capabilities.
For example, Dassault Systèmes defines the notion of a \textit{virtual twin} as ``\textit{the representation of the full lifecycle, behavior and evolution of a product or system, starting with a 3D model that captures the shape, dimensions and properties of a physical object}.''\footnote{\url{https://www.3ds.com/virtual-twin}} Dassault Systèmes achieves this through advanced 3D experience, enabled by their 3DEXPERIENCE platform.\footnote{\url{https://www.3ds.com/3dexperience/}}

With the uptick in the adoption of digital twins, their general typology has experienced a noticeable shift. Reports both from industry~\cite{muctadir2024current} and academia~\cite{dalibor2022cross} hint at an overall lack of understanding of digital twins and the surrounding ecosystems. Today, digital twins are used as an umbrella term that covers real digital twins and some limited variants, too, e.g., digital shadows and even digital models. While there is often no point in arguing against such umbrella concepts, we provide a sufficiently clear, fit-for-purpose definition of digital twins.

\phantom{}

\begin{defframe}
    A digital twin is a real-time, sufficiently realistic computational reflection of a system with the ability to control it.
\end{defframe}

This high-level definition turns out to be adequate for the vast majority of situations, and there are some notable properties it implies and that merit discussion.

\begin{description}
    \item[Computational reflection] is provided by a model that is continuously updated with data from the physical twin, to reflect the prevalent state of the twinned system. This model is used by computer machinery for the automated reasoning about the physical system's state; and by humans to observe, inspect, and operate physical systems.

    \item[The model] is an abstraction of reality, capturing only the essential properties of the system~\cite{stachowiak1973allgemeine,minsky1965matter}, and it is valid under a limited set of conditions~\cite{zeigler2018theory}.
    
    \item[The model] is digital. This is contrasted with the actual system, which is typically of physical nature. The model is, therefore, amenable to computer-automated analysis and simulation.
\end{description}

\subsubsection*{Examples of digital twin systems} To illustrate the above definition, here are some examples of systems that rely on digital twins.

\begin{description}
    \item[Robotic assembly system for smart manufacturing,] in which a manufacturing system is the physical twin and it is observed and controlled through a cloud-based digital twin~\cite{touhid2023building}. The digital twin contains models of the geometry of robot arms, models of kinematics to understand and control its movement and states (e.g., safe states, unstable states, etc). Such solutions constitute the foundations of Industry 4.0 and Industry 5.0.
    
    \item[Smart building,] in which the building is the physical twin and it is monitored and controlled for optimal temperature~\cite{elmokhtari2022development}. In the digital twin, 3D models and models of physics aid automated decision-making, integrated into a unified building information model. Such solutions constitute the foundations of Construction 4.0.
    
    \item[Controlled environment agriculture,] in which a smart agriculture enclosure is the physical twins and it is monitored and controlled for optimal crop growth conditions, including temperature, humidity, and lighting conditions~\cite{david2023digital}. Discrete event models may capture states of machines (e.g., on/off), and continuous (e.g., Simulink-based) energy models may capture the energy states of plants. Equipment (e.g., irrigation and air conditioning) are controlled based on insights from high-performance simulators.
\end{description}

\subsubsection*{Examples of systems that are not digital twins,} at least not in the strict sense.

\begin{description}
    \item[A monitoring software] is not a digital twin in the strict sense. A monitoring software has no control over the physical system, and typically, monitoring software has no central computational model to reflect the system's state.
    
    \item[Closed-loop control] is not a digital twin. In some research communities, closed-loop control is considered a form of a digital twin. This is not entirely correct. While the control element is clearly there, a controller does not facilitate the computational reflection of the system.
    
    \item[Models@run.time] are not digital twins. Models@run.time~\cite{blair2009models} support computational reflection of physical systems but are not capable of controlling them.
\end{description}

\section{The case for digital twins in smart grids}\label{sec:dt-in-smart-grid}

Power grids are a critical infrastructure to modern society. They enable nowadays' digitalized world and economy, and directly affect our daily lives. Providing safe, stable, and sufficient electricity is essential for sustainable development worldwide. However, the power industry is facing serious challenges: growing energy demand, limited resources, and the looming threat of climate change render traditional power generation and distribution practices insufficient~\cite{mchirgui2024applications}. The growing electricity demand from electric vehicles and data centers requires more flexible and responsive grid management that supports a sustainable energy transition. However, traditional centralized power generation cannot efficiently integrate variable renewable energy sources essential for carbon emission reduction, such as solar and wind energy. These challenges create an urgent need to modernize power infrastructures.
Smart grids tackle these challenges by using two-way flows of electricity and information to create a widely distributed automated energy delivery network~\cite{fang2011smart}. In smart grids, electricity can flow bi-directionally, and information flows multi-directionally (Figure~\ref{fig:dt-powergrid}). This enables, for example, transforming consumers into \textit{prosumers} (producer-consumers) who generate power through small-scale power generators, such as rooftop solar panels and small wind turbines, and feed excess electricity back to the grid~\cite{sarwar2016review}. Prosumers continuously share consumption and generation data with grid operators, who respond with dynamic pricing signals and demand management requests. This creates a dynamic system where smart grids constantly adjust power supply and distribution based on real-time data from thousands of distributed energy sources.

Apart from being highly dynamic systems, the topology and organizational structure of smart grids changes rapidly too, for example, due to prosumers adding and removing distributed generation sources, system overloads, and occasional outages~\cite{pan2020digital}. 
Meanwhile, advanced technologies such as Internet of things (IoT) sensors for real-time monitoring, advanced metering infrastructure (AMI), large-scale battery-energy-storage systems, and vehicle-to-grid (V2G) chargers are being introduced to the power sector, making it more complex. 
These new smart technologies greatly enrich computing and communication capabilities and make power management more efficient, reliable, and sustainable. However, they create new operational challenges that cannot be addressed through physical infrastructure alone~\cite{jafari2023review}.

Together, dynamic prosumer behavior and advanced technological complexity of smart grids create a control problem: real-time load balancing must now reconcile thousands of bidirectional prosumers whose injections rise and fall with weather, tariffs, and local storage schedules.

Digital twins offer a powerful solution to address these challenges across the entire lifecycle of smart grid systems. The digital twin keeps a synchronized virtual replica of each asset and of the grid as a whole, so that planners and cyber-defenders can all act on shared, up-to-date data. This replica underpins day-to-day monitoring, optimization, security analytics and predictive maintenance~\cite{das2024advancements}, and it becomes a real-time decision-support tool when contingencies strike, such as equipment failure, and coordinated cyber-attacks~\cite{zheng2022smart}. Furthermore, digital twins enable real-time coordination of thousands of prosumers by continuously modeling their bidirectional power flows and predicting their behavior based on weather patterns, tariff structures, and storage schedules, creating a closed-loop control system that automatically adjusts physical grid operations to maintain optimal performance. When smart grids and digital twins work together, energy providers can achieve both operational efficiency and environmental sustainability goals through optimized renewable energy integration~\cite{irfan2023digital,sleiti2022digital}.

In the following, we look at how digital twins play a crucial role in four lifecycle phases of a smart grid: design, testing \& verification, operation, and maintenance \& improvement. Through real-world examples and use cases, we discuss how the unique capabilities of digital twins enable new ways to enhance grid reliability, resilience, and performance.

\begin{figure}
    \centering
    \includegraphics[width=0.75\linewidth]{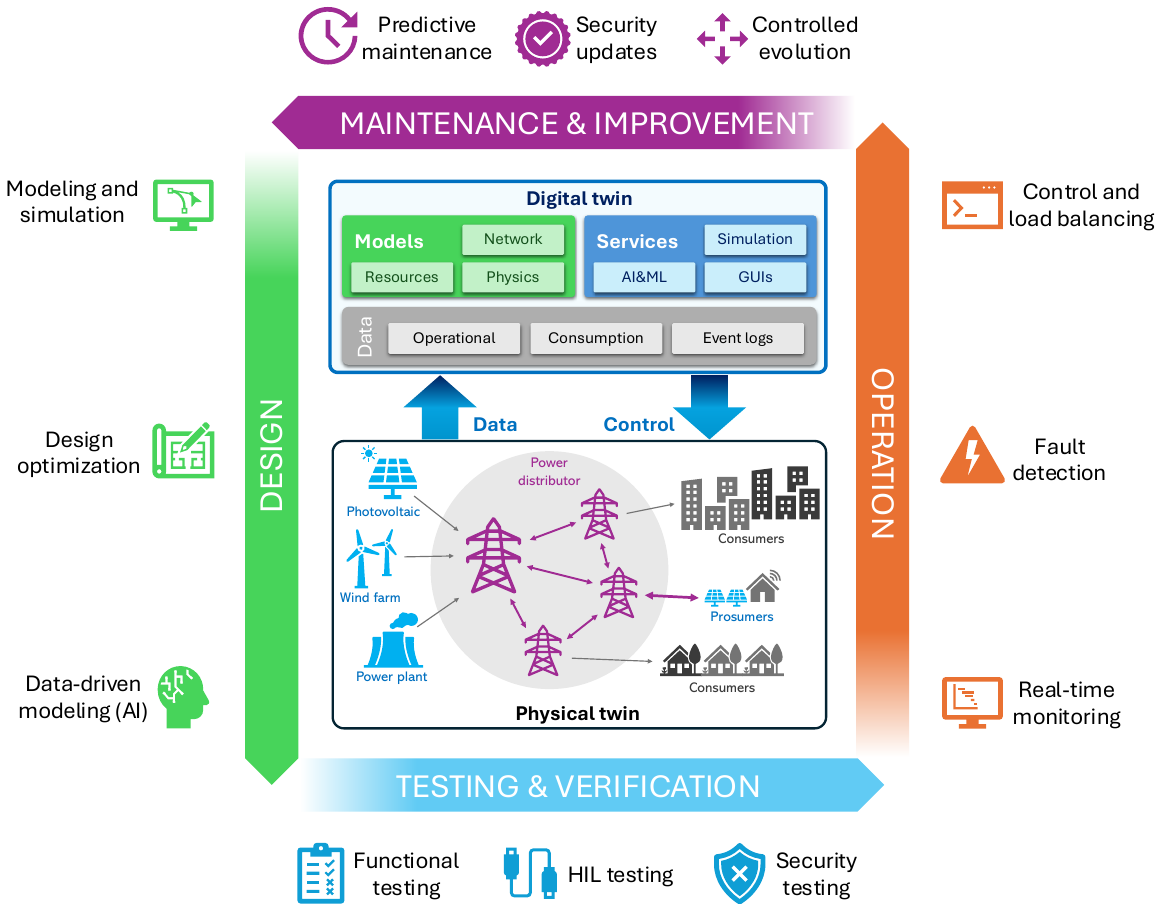}
    \caption{Key digital twin responsibilities along the lifecycle of smart grids}
    \label{fig:dt-grid-lifecycle}
\end{figure}

\subsection{Design}
In the design of smart grid infrastructures, engineers build various analytical and simulation models of the physical system. These models constitute the primary artifacts later used in a digital twin. Digital twins in the design phase can be used, for example, to simulate the impact of adding a new solar farm or wind park to the grid before construction, explore the associated costs and risks, and visualize recommended design decisions for key stakeholders.

\subsubsection*{Modeling and simulation}
During the design phase, digital twins serve as foundational tools built on two core capabilities: comprehensive modeling and simulation of smart grid systems, and the connected nature between digital and physical twins that enables continuous synchronization. These digital representations allow engineers to create virtual smart grid systems, addressing various aspects of design complexity through different modeling approaches. Several distinct modeling approaches emerge to address specific aspects of smart grid design complexity.
One comprehensive approach involves modeling both physical and cyber layers simultaneously. For example, the ANGEL digital twin employs such a dual-layer design methodology, in which integrated models represent both the physical grid infrastructure and its cyber-security framework~\cite{danilczyk2019angel}. This design approach ensures that security considerations are embedded into the system architecture, enabling the digital twin to validate data integrity and evaluate grid performance within a unified modeling framework.
For real-time applications, probabilistic modeling offers advantages. \citet{milton2020controller} propose an approach for online diagnostic analysis of power electronic converters using real-time, probabilistic digital twinning. The approach involves developing real-time, probabilistic simulation models with stochastic variables and embedding them into power converter controllers running on field programmable gate array (FPGA) computing devices—specialized hardware that can be reconfigured for specific computational tasks.

\subsubsection*{Design optimization}
Digital twins also support design optimization by leveraging operational insights through a feedback loop from operational monitoring back into the design phase.
For example, rotor speed data collected from an operating wind turbine can be analyzed to determine the most efficient speed for power generation. These resulting insights can be fed back into the design phase to inform future designs or modifications of wind turbine systems~\cite{wang2021recent}.

At a broader system level, digital twins enable strategic infrastructure planning and optimization. Advanced analytics applied to the digital twin can help discover ideal load distribution patterns, reduce peak load pressures, and recommend modifications to improve system resilience and efficiency~\cite{jafari2023review}.

\subsubsection*{Data- and AI-driven modeling}
Building upon these foundational modeling capabilities, advanced techniques like artificial intelligence (AI) can further enhance digital twin functionalities in smart grid design. The convergence of AI, IoT, and cyber–physical systems with smart grid infrastructure enables the continuous gathering, processing, and analysis of real-time data to accurately assess current and future states~\cite{jafari2023review}. One particularly important application of AI-enhanced digital twins is climate resilience planning. Climate resilience modeling integrates physical hazard assessment with data-driven capabilities. An innovative approach introduces modeling and predicting the performance of electric power networks (EPNs) when exposed to extreme weather events~\cite{braik2023novel}. The suggested digital twin framework integrates physical models, such as hazard and vulnerability assessments, with data-centric methodologies, specifically, a dynamic Bayesian network (DBN), to create a high-fidelity hybrid model that can receive updates in near-real time through sensor data.

\subsection{Testing \& Verification}
Once comprehensive digital twin models are established during the design phase, they provide the foundation for rigorous testing and validation before system deployment.
Digital twins provide a safe and cost-efficient environment for testing extreme scenarios that would be impossible or too risky to perform on actual grid infrastructure~\cite{mansour2023applications}. Key examples of such scenarios pertain to the safety of the grid, e.g., testing the impact of extreme weather conditions or electromagnetic pulse (EMP) attacks; as well as the security of the grid, e.g., testing cybersecurity attacks through a coordinated false data injection attack on multiple substations.

\subsubsection*{Functional testing}
By creating a dynamic digital replica of the physical grid, operators can simulate various load scenarios in controlled environments before they occur in the real world. This simulation capability allows teams to test how the system responds to potential imbalances, identify weak points, and verify that protection systems respond appropriately.
For example, engineers can simulate scenarios with rapid load fluctuations or unexpected demand spikes to verify system stability~\cite{mchirgui2024applications}. These tests help validate the effectiveness of existing control systems and protection mechanisms, ensuring they perform as designed under stress conditions. This capability allows grid operators to explore worst-case scenarios and boundary conditions that would otherwise remain theoretical, generating practical knowledge that helps prevent costly failures and outages. By testing various dangerous scenarios on the digital copy first, engineers can predict potential problems before they ever happen in the real world.
A practical example of this safe testing environment is when digital twins are used for wind turbines~\cite{wang2021recent}. Engineers can create a physics-based model---a digital twin---of an actual wind turbine to test how it would perform in extreme conditions. For instance, they can simulate what happens when the turbine faces dangerously high wind speeds or when it operates below its normal cut-off speed.

\subsubsection*{Hardware-in-the-loop (HIL) testing}
Digital twins allow for advanced hardware-in-the-loop testing of critical power grid components too~\cite{nguyen2022digital}.
Integrating experimental hardware components into the actual grid would be too risky. Digital twins can be used, for example, in the pre-integration phase to assess the impact of hardware elements---for example, relays, controllers, and storages---to be deployed on the grid. Digital twins, in such a setup, simulates the behavior of the overall grid, while Power-Hardware-in-the-Loop (PHIL) components render test scenarios more realistic. This allows for detecting rare edge cases (so-called ``black swan events'') in a simulated grid that pure in-silico models may miss.

\subsubsection*{Security testing}
Digital twins also allow for security testing smart grids. Traditional security testing on active power grids is problematic because it can disrupt service to thousands of customers. With digital twins, security experts can accurately model how the physical grid functions and run comprehensive security tests without causing any service disruptions. This approach establishes a common framework for developing standardized models that allow for continuous and thorough security testing~\cite{atalay2020digital}. Security teams can repeatedly test for vulnerabilities, update their testing procedures as new threats emerge, and ensure that the grid remains protected, all without affecting the day-to-day operation of the actual power system. Another critical application is testing protection systems, the safeguards designed to prevent cascading failures in the grid~\cite{kuber2022virtual}. A digital twin of a relay contains all the same protection functions and algorithms as the physical device. 
This virtual approach goes beyond just verifying individual components; it allows engineers to validate how the entire protection system works together without needing physical access to substations. This saves time, reduces costs, and improves safety for maintenance crews.

\subsection{Operation}
During the operational phase, digital twins are live-synchronized systems that provide real-time insights and control capabilities. These virtual replicas continuously stream data and analytics to support maintenance decisions, user interactions, and immediate fault detection across the entire grid infrastructure. A pertinent example is real-time voltage regulation and congestion management. By maintaining a synchronized virtual replica of the grid, the digital twin continuously monitors node voltages and line loadings. When demand peaks or renewable generation fluctuates, the twin can simulate corrective actions such as reactive power dispatch from distributed energy resources or reconfiguration of network topology. These simulations allow operators to identify the most effective intervention in real time, preventing voltage collapse or line overloading.

\subsubsection*{Real-time monitoring, control, and load balancing}
Digital Twins provide operators with comprehensive, up-to-date information about all grid components and systems. This continuous monitoring capability extends across various grid assets, from individual components to entire subsystems, enabling precise operational control and decision-making~\cite{mchirgui2024applications}.
For wind turbines~\cite{wang2021recent}, digital twins collect comprehensive operational data including output power, wind conditions, temperature, electrical currents, and mechanical parameters. This data can be used to analyze rotor speed performance and ensure safe operation within established limits. For energy storage systems, digital twins are used to visualize, process, and assess the performance in the virtual environment. This technology is used to determine the scheduling programs for the operational process~\cite{steindl2021semantic}. 

\subsubsection*{Real-time fault detection}
Digital twins can process real-time data from both cyber and physical domains to detect anomalies, and their multidimensional characteristics enable the real-time detection of faults that improve overall grid reliability and stability~\cite{khan2023digital}.
In a typical case, researchers implemented a digital twin for 110 kV oil-filled transformers that integrated multiple diagnostic sensors, including ultraviolet monitoring, thermal imaging, and discharge measurements to create a comprehensive digital model~\cite{khalyasmaa2020digital}. This multi-sensor digital twin provided instant feedback on transformer condition and detected performance degradation over time, enabling operators to identify potential issues before they escalated into costly failures. For distributed energy resources, researchers develop neural network-based fault detection systems based on a digital twin. These systems use two-stage machine learning tools to process massive data streams from smart meters, identify potential fault locations, and reduce the complexity of monitoring distributed grid components~\cite{tzanis2020hybrid}.

\subsubsection*{Cyber-security and real-time threat detection}
Digital twins enhance cyber-security through real-time data analysis capabilities that improve cyber-attack detection performance while reducing response times~\cite{khan2023digital}.
Advanced security frameworks like ANGEL demonstrate sophisticated protection capabilities by modeling both cyber and physical grid layers. The system provides real-time data visualization and health assessment across various operating conditions, enabling evaluation of both current system status and potential security threats. Using physics-based modeling combined with machine learning algorithms, ANGEL create comprehensive defense mechanisms that reduce grid vulnerabilities and accelerate threat detection~\cite{danilczyk2019angel}. For complex grids like China's ultra-high voltage systems, digital twins provide essential decision support with sub-second response times. These real-time analysis systems handle unprecedented operational complexity while maintaining rapid responsiveness for critical security decisions~\cite{zhou2020real}.

\subsubsection*{Customer engagement and demand response}
Digital Twins transform customer participation in grid operations by providing unprecedented visibility into energy usage patterns. Instead of waiting for monthly bills, customers receive real-time information about their electricity consumption, timing, and costs.
Smart meters serve as the primary interface for this enhanced engagement~\cite{das2024advancements}, enabling customers to understand their energy usage patterns and make informed decisions. When combined with real-time pricing, this visibility allows customers to reduce costs by shifting consumption away from peak-price periods, contributing to overall grid efficiency.

\subsection{Maintenance \& Improvement}
The wealth of operational data collected through digital twin monitoring creates opportunities beyond day-to-day operations. As digital twins collect operational data over time, they become increasingly valuable for planning future upgrades and optimizations. The historical information captured by digital twins provides insights that would be impossible to gather from the physical system alone, creating a foundation for continuous improvement~\cite{jafari2023review}. For example, by analyzing vibration and temperature data in a digital twin, operators can predict a bearing failure in a grid-scale transformer weeks in advance, allowing for planned maintenance instead of an emergency outage.

\subsubsection*{Predictive maintenance}

By predicting maintenance needs based on actual asset conditions rather than predetermined schedules, digital twin technologies reduce unnecessary maintenance, extend equipment life, and ensure optimal performance~\cite{coppolino2023building}. By deploying Industrial Internet of Things (IIoT) networks and computational resources at the edge of the system, it becomes possible to collect and process both historical and real-time data directly at the source. This enables data-driven machine learning algorithms with the support of digital twin analytics to generate optimized maintenance schedules for each subsystem~\cite{khan2023digital}. 
The dynamic nature of digital twins allows operators to model the grid's evolution over extended periods, simulating how load patterns might change over months and years~\cite{jafari2023review}. This long-term perspective allows utilities to predict and mitigate potential imbalances that might emerge as consumption patterns change, renewable generation increases, and new technologies like electric vehicles become more prevalent. Additionally, predicting the remaining useful life (RUL) of components in the smart grid becomes more accurate and efficient. Digital twin can forecast when components are likely to fail, allowing operators to schedule maintenance before problems occur~\cite{aizpurua2017determining}. This predictive approach helps avoid unexpected failures and costly service disruptions.

\subsubsection*{Security updates}
Unlike traditional security approaches that remain static until breaches occur, digital twins enable an adaptive security posture that evolves based on emerging threats. \citet{coppolino2023building} implement a digital twin in a real-world case study on an operational smart grid in Kropa, Slovenia, where the twin provided a comprehensive view of the grid's health and enabling quick detection and communication of potential threats. As threats evolve, the digital twin captures new attack patterns and automatically updates its security protocols, creating a continuously improving security posture that stays ahead of emerging risks.

\subsubsection*{Controlled evolution}
Digital twins enable controlled system evolution through operational optimization. Wind turbine operations provide an illustration~\cite{wang2021recent}. Wind turbine blade pitch (the angle of the blades relative to the wind) significantly affects both energy production and equipment longevity, and finding the perfect pitch angle for varying wind conditions has traditionally been challenging. With digital twins, an optimization program can use the digital twin to test hundreds of pitch angle settings for various wind speeds, directions, and turbine locations in order to find the best configuration. Then, such optimization can be used to close the control loop that controls the wind turbine's blade rotation.
As power grids incorporate more diverse energy sources, digital twins provide a platform for continuous optimization across these hybrid resources. For example, the system can learn seasonal patterns in renewable generation, gradually adjusting control algorithms to maximize renewable utilization, reduce operational costs while maintaining grid stability~\cite{khan2023digital}. This knowledge can then inform future infrastructure investments.

\section{Engineering digital twins}\label{sec:sw}

The tight coupling of digital twins with physical systems merits special considerations in their development. Here, we review some of the key software organization and architectural guidelines that are used in digital twin engineering.

\subsection{The ISO 23247 reference architecture}

Despite the wide adoption of digital twins, their engineering methods are still in the early stages of maturity. This, in turn, gives rise to some pressing software engineering problems, such as the relative scarcity of architectural guidelines and standards. Especially in regulated domains---such as smart grids---software certification is a typical need~\cite{song2019review}. However, in the absence of standards, certifying software is a costly endeavor.

\begin{figure}
    \centering
    \includegraphics[width=0.9\linewidth]{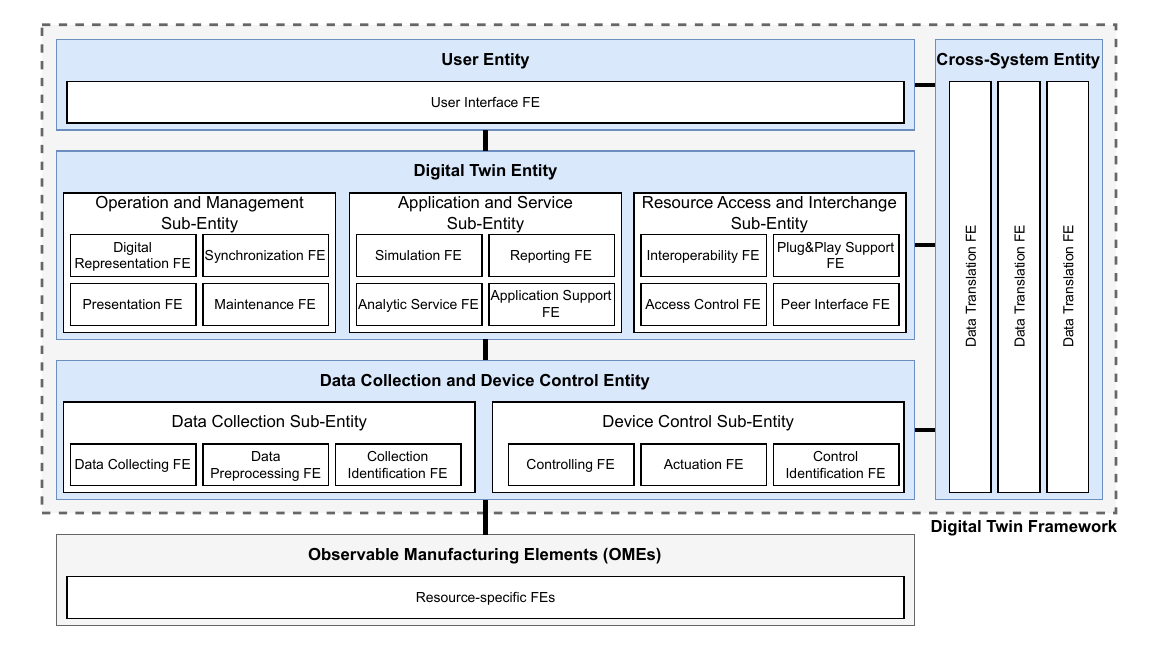}
    \caption{Functional view of the ISO 23247 digital twin reference architecture. (Reproduced from~\cite{shao2021use}.)}
    \label{fig:iso-arch}
\end{figure}

The \textit{ISO 23247 - Digital Twin Manufacturing Framework} standard is a relatively new one, specifically tailored for digital twins.\footnote{\url{https://www.iso.org/standard/75066.html}} Published in 2001 by the International Organization for Standardization (ISO), the standard defines a development framework for digital twins, particularly in the manufacturing domain. It consists of four parts: Part 1 -- Overview and general principles; Part 2 -- Reference architecture; Part 3 -- Digital representation of manufacturing elements; Part 4 -- Information exchange. 
For software architecture purposes, Part 2 provides a detailed, layered reference architecture.\footnote{\url{https://www.iso.org/standard/78743.html}} As shown in Figure~\ref{fig:iso-arch}, the reference architecture consists of domains, refined into domain entities, sub-entities, and, eventually, functional entities (FEs). A functional entity is a unit of coherent functionality ISO-compliant digital twins shall implement. For example, the \textit{Simulation FE} predicts the behavior of physical entities, the \textit{Analytic service FE} manages and analyses data collected from physical entities, and the result of simulations, the \textit{Reporting FE} generates reports about production results, simulation predictions,
and the results of data analytics, and the \textit{Application support FE} provides services for implementing applications such as predictive and reactive maintenance, open and closed loop applications. These four FEs together constitute the \textit{Application and Service Sub-Entity} whose role is to provide functionality related to applications and services. This sub-entity is situated within the \textit{Digital twin entity}, which is the (set of) system(s) providing functionalities for digital twins, such as realization, management, synchronization, and simulation. The reader is referred to \citet{shao2021use} and \citet{shao2023analysis} for detailed elaboration on the ISO 23247 standard.

Standardization of and for digital twins is subject to active research and development. Since ISO 23247 is currently the only digital twin-focused standard, its limitations are often on display in domains outside of manufacturing. For example, the reference architecture does not provide a specific functional entity for enhanced data storage, despite acknowledging the need for data exchange via databases or cloud systems~\cite{ferko2023standardisation}. Advanced analytics and deep-learning FEs are also missing~\cite{kang2024edge}. The need for such FEs has been identified, e.g., in digital twin-enabled AI simulation~\cite{liu2025ai}. The standard does not address verification, validation, and uncertainty quantification (VVUQ) tasks, which are fundamental in regulated domains~\cite{shao2024digital}. Finally, while there exists an \textit{Interoperability Support FE} in the standard, it is seldom implemented in practical applications~\cite{ferko2023standardisation}, leaving digital twin aggregation and composition efforts without actionable guidelines~\cite{adesanya2024systems}.

Fortunately, there are some important ongoing standardization efforts for digital twins, including two new parts to the ISO 23247 standard -- Part 5: digital thread for digital twins, and Part 6: digital twin composition~\cite{shao2024manufacturing}. The joint technical committee of the International Organization for Standardization (ISO) and the International Electrotechnical Commission (IEC) -- ISO/IEC JTC 1 -- is currently working on refining definitions and concept related to digital twins, as well as developing maturity models and reference architectures. The Digital Twin Consortium of the Object Management Group (OMG) is another consortium that works on standardization, with a focus on cross-domain industry standards.

In the interim, more purpose-specific standards can be used to augment digital twins, particularly for data management, data security, and information modeling. For example, the OPC Unified Architecture (OPC UA)\footnote{\url{https://opcfoundation.org/developer-tools/documents/?type=Specification}} is a frequently used standard in complex digital twin applications where integration is a concern~\cite{adesanya2024systems}. It defines a standardized framework for reliable and secure communication among diverse subsystems.

The reader is referred to \citet{shao2024manufacturing}, \citet{ferko2023standardisation}, and \citet{david2025interoperability} fo further pointers on standardization.

\subsection{Reference Architectural Model Industrie 4.0 and the Asset Administration Shell}

\begin{figure}
    \centering
    \includegraphics[width=0.8\linewidth]{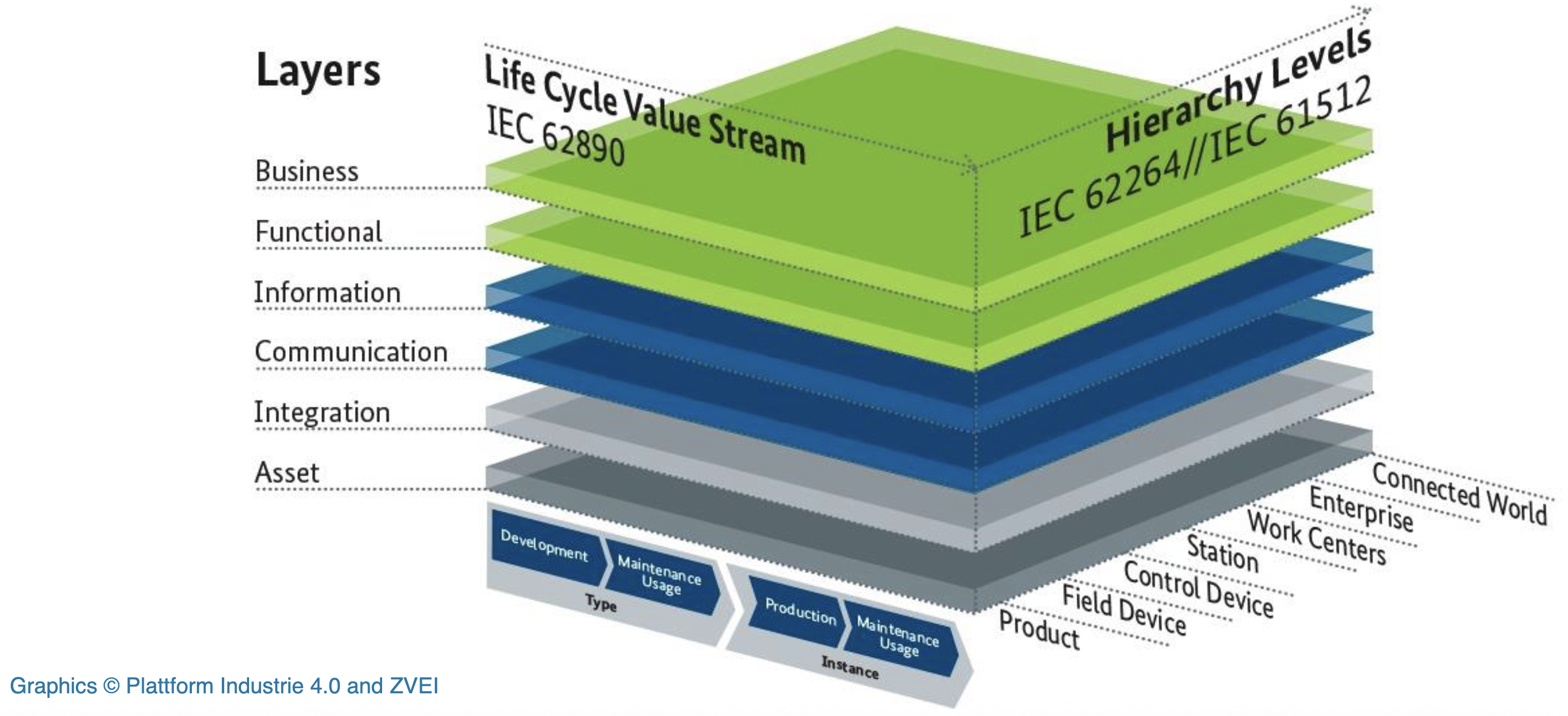}
    \caption{Overview of the RAMI 4.0 reference model~\cite{rami40}}
    \label{fig:rami40}
\end{figure}

Another important reference model is the Reference Architectural Model Industrie 4.0 (RAMI 4.0). Compared to the ISO 23247 standard, RAMI 4.0 defines higher-level guidelines, focusing on how to situate Industry 4.0 systems, such as digital twins in the overall corporate information technology (IT) infrastructure. As shown in Figure~\ref{fig:rami40}, RAMI 4.0 defines the following three dimensions to approach Industry 4.0 system development in a structured fashion.

\begin{description}
    \item[Hierarchy levels] concern the value creation systems that create the product, ranging from the \textit{Product} itself through work \textit{Station}s to the external, \textit{Connected world}. This hierarchy is compliant with the IEC 62264 (Enterprise-control system integration)\footnote{\url{https://www.iso.org/standard/57308.html}} and IEC 61512 (Batch control)\footnote{\url{https://webstore.iec.ch/en/publication/107554}} standards.
    \item[Life cycle value stream] allows for describing the lifecycle of the product from early conceptualization to retirement. The lifecycle model is compliant with the IEC 62890 (Industrial-process measurement, control and automation - Life-cycle-management for systems and components)\footnote{\url{https://webstore.iec.ch/en/publication/30583}} standard, which establishes basic principles for lifecycle management of systems and components used for industrial-process measurement, control and automation.
    \item[Layers] are IT layers, specifically, and allow for describing products and systems in terms of, e.g., physical components (\textit{Asset}), the data they operate with (\textit{Information}), and the functionality such systems provide (\textit{Functional}). It is easy to see how these layers correspond to the concept of digital twin.
\end{description}

The three-dimensional model of RAMI 4.0 establishes a common vocabulary for stakeholders and fosters a better understanding of the role and implementation of digital twins in an Industry 4.0 context. The key benefit of RAMI 4.0 is that it provides a unified framework across three orthogonal dimensions, within which processes, component hierarchies and IT concerns are properly aligned. By relying on standards in each dimensions, the RAMI 4.0 framework actively supports independence from vendors, products, and technologies.

Subject matter experts and stakeholders typically interact with the RAMI 4.0 framework through domain-specific architectural views and viewpoints. These viewpoints slice the three dimensions in a way that is the most useful for their purposes. For example, a business process viewpoint might be interested in business--functional--informational layers in the station--work center--enterprise hierarchies, throughout the entire lifecycle. Or, an integration viewpoint might be interested in the communication--integration layers in the development--maintenance phases of the product.
In general, views are viewpoints applied to a specific problem, model, or architecture. A more formal definition is given by the ISO/IEC/IEEE 42010:2022 (Software, systems and enterprise --- Architecture description) standard\footnote{\url{https://www.iso.org/standard/74393.html}}, which defines an architecture viewpoint as a set of conventions for the creation, interpretation and use of an architecture view; and an architecture view as the information part comprising portion of an architecture description. We remark here that in some communities and industries, ``view specification'' and ``viewpoint'' are synonymous.

The RAMI 4.0 framework supports digital twin technology in numerous ways. First, it defines a high-level, standardized model for digital twins to organize the key components of digital twins (physical assets, digital representations) along with their integration points (data and control) in a clear lifecycle model. Digital twins are typically realized at the upper layers of RAMI 4.0.
While the framework allows for detailed modeling and operationalization of digital twins, its richness also increases the complexity of implementing digital twins.  To alleviate this issue, the Asset Administration Shell (AAS) has been developed by the German Electrical and Electronic Manufacturers' Association for the standardized digital representation of assets of the Reference Architecture Model Industrie 4.0 (RAMI4.0)~\cite{oakes2023digital}. The AAS provides a machine-readable, device-independent, hierarchical standard language (metamodel) for describing the properties of assets. As such, the AAS is a key enabler of digital twin technology within the RAMI 4.0 framework.

\subsection{Open challenges in implementing digital twins for the smart grid}

As technological capabilities increase, the industry is increasingly recognizing the disruptive potential of digital twin technologies. However, the development of digital twins for smart grids is hindered by several challenges due to the complex nature of the power infrastructure.
Here, we review some of the most pressing challenges.

\begin{description}
    \item[Data infrastructure and management] The foundation of any successful smart grid digital twin implementation rests on a robust technical infrastructure capable of handling massive data volumes from smart meters, Supervisory Control and Data Acquisition (SCADA) systems, and grid sensors. Current IT systems often prove inadequate for supporting the intensive real-time data analysis demands required for smart grid operations, creating a fundamental barrier to implementation~\cite{jafari2023review}. Also, smart grid digital twin systems generate enormous amounts of data from distributed energy resources, smart meters, and grid monitoring devices, which must be processed and analyzed in real-time~\cite{mchirgui2024applications}. To address these infrastructure limitations, organizations can invest in high-performance computing solutions, including advanced GPUs and cloud-based services from providers like Amazon, Microsoft, and Google~\cite{jafari2023review}. Edge computing emerges as a particularly valuable approach, reducing data transmission delays while increasing bandwidth efficiency~\cite{ezeugwa2024evaluating}. For effective data management, implementing distributed storage solutions becomes essential, e.g., using Hadoop Distributed File System (HDFS) and cloud-based platforms like Amazon S3, Microsoft Azure Blob Storage, or Google Cloud Storage~\cite{mohammed2023digital}. These technologies provide the scalable and reliable storage and retrieval capabilities necessary for handling the vast datasets generated by digital twin applications.
    
    In smart grids, an additional challenge lies in integrating heterogeneous legacy systems, such as legacy SCADA architectures, analog meters, and proprietary communication protocols with modern IoT-based data streams. Without proper middleware, data adapters, and semantic alignment, these legacy systems may create information silos or introduce latency and inconsistency into the digital twin. Therefore, interoperability standards and data integration frameworks---such as DMTF’s Common Information Model (CIM)\footnote{\url{https://www.dmtf.org/standards/cim}} and the earlier mentioned OPC UA---become critical for achieving a unified, coherent view of the grid across both old and new infrastructure.

    \item[Connectivity and real-time processing] Smart grid digital twins depend on seamless connectivity between physical grid infrastructure and their virtual counterparts, yet this remains one of the most challenging aspects for utility operators. Despite advances in smart grid communication technologies, significant connectivity issues persist in substation networks, smart meter communications, and distribution automation systems,  including software errors, power outages, and missing data that can disrupt critical grid monitoring capabilities~\cite{jafari2023review, das2024advancements}. In addition, the latency in data collection from IoT devices such as smart meters and grid sensors limits the efficiency of digital twins~\cite{almasan2022digital}. Given that the quality and timeliness of IoT data streams fundamentally determine digital twin performance, smart grid operators should prioritize robust communication protocols and implement backup data collection methods to ensure continuous, reliable connections between physical and digital twins~\cite{ebrahimi2019challenges}.

    To meet these demands, grid providers often adopt hybrid communication architectures that combine multiple technologies, e.g., fiber-optic links for critical backbone infrastructure, 5G and cellular networks for distributed assets, and radio-frequency or low-power wide-area networks for remote sensors in areas that are hard to reach. Such a hybrid approach improves coverage and redundancy; but at the same time, it introduces latency management challenges, as data from heterogeneous networks needs to be synchronized, time-stamped, and fused into a coherent stream for real-time decision-making. These challenges can be addressed, e.g., through intelligent edge processing and time-sensitive networking (TSN)~\cite{zhang2024time-sensitive} which may help minimize delays.

    \item[Security] The extensive data exchange inherent in smart grid digital twin systems creates substantial security vulnerabilities that pose risks to power infrastructure \cite{jafari2023review}. These systems collect and process sensitive grid operational data from multiple sources, including IoT devices, SCADA systems, and grid sensors, making them high-value targets for cyberattacks. The interconnected nature of smart grid networks amplifies security risks, as vulnerabilities in one component can potentially compromise the entire system \cite{casola2023opportunities}. Ensuring data security requires implementing multiple layers of protection, starting with advanced encryption methods, where blockchain technology provides a particularly effective solution through its decentralized and tamper-resistant architecture \cite{ebrahimi2019challenges}. Organizations can also adopt privacy-enhancing technologies and systematic data governance practices to minimize exposure risks. These protective measures include strategic data minimization to limit the volume of sensitive information collected, robust anonymization techniques to obscure individual identities, and comprehensive identity protection protocols to prevent unauthorized access and potential data breaches~\cite{botin2022digital}.
    
    Ensuring security requires implementing multiple layers of protection through advanced encryption methods. Privacy-enhancing technologies and systematic data governance practices help further minimize exposure risks. Beyond simple data breaches, an exposed digital twin may influence the grid in adverse and even catastrophic ways. Successful cyberattacks may potentially propagate through the physical grid, causing service disruptions or equipment damage. Thus, a digital twin of a smart grid must be treated as a critical attack surface and protected with the same rigor as, e.g., SCADA and control systems.

\end{description}

\section{Advanced Topics in Digital Twin Technology}\label{sec:advanced-topics}

There are some advanced topics in digital twin technology that merit further discussion in relation to smart grids.

\subsection{Digital Twins and AI}

The success of digital twin technology hinges on the quality and fidelity of the models that capture essential properties of the physical twin. Complex physical systems, however are not trivial to model manually. Smart grids are pertinent examples of such systems, as they require modeling of physics, power, production-consumption dynamics, environmental conditions, as well as fault modes and security concerns. AI can alleviate the complexity of modeling in digital twins in at least two ways.

\subsubsection*{Surrogate models}

Surrogate models are simplified or reduced models that are used in place of more complex models. Surrogate models approximate the behavior of more complex, expensive, or computationally intensive models, but often, these approximations are sufficient and the reduced computation time and costs compensate for the lost precision. AI models are prime examples of surrogate models in modern systems engineering. Instead of manually modeling a complex system (e.g., the physics of a smart grid), a surrogate AI model can be trained on operational data and used in a digital twin to replace resource-intensive numeric calculations with approximations the AI model has been trained for. Clearly, a surrogate AI model will provide reliable results only for the cases it has been trained for, while a general numeric model is more generally applicable. Thus, a surrogate AI model finds trade-offs between generality and speed, too. Some of the typical examples of AI models used in modern digital twins are the following.

\begin{description}
    \item[Deep neural networks (DNNs)] can learn complex, hierarchical patterns from data with sufficiently high number of dimensions. DNNs are structured into multiple hidden layers, which progressively transform data into increasingly complex structures. DNNs are often used, e.g., for energy forecasting in a smart building, a DNNs might take environmental conditions (temperature, humidity) and occupancy information (hour of day) as input, detects correlations in the first layer (e.g., high temperature $\rightarrow$ more AC usage), combines these correlations in the second layer (e.g., high temperature outside + the building is crowded $\rightarrow$ maximum AC usage and energy consumption), and output the energy prediction in the output layer.

    \item[Physics-informed neural networks (PINNs)] improve over DNNs by integrating physical laws (e.g., differential equations) directly into the learning process. DNNs may produce output that is not possible in the real world. For example, a DNN trained to predict voltage in a smart grid might give plausible outputs, but violate power flow equations during unusual demand spikes. A PINN, on the other hand, is trained to respect the underlying physics, ensuring predictions remain realistic. PINNs achieve this by building laws of physics directly into the loss function during training. A by-product of these physics-based hints during training is the reduced data requirements of training PINNs.
        
    \item[Gaussian Process Regression (GPR)] is a non-parametric Bayesian approach for interpolation in Gaussian processes. GPR is a probabilistic machine learning method that predicts functions based on observed data, wind flows over a turbine. GPR is uncertainty-aware, i.e., the variance of predictions is known. GPR also works well with limited datasets. These two properties are the strengths of GPR over DNNs.
\end{description}

The choice between a manually-built model and an AI surrogate depends on the objective of the digital twin. For safety-critical applications, such as real-time grid stability control, fault detection, or protection coordination, manually-built models are preferred because they are transparent and easier to verify and certify. In contrast, AI surrogate models act as black-boxes and are harder to verify and certify and as a consequence, less suitable for safety-critical applications. However, AI surrogates are excellent for tasks where rapid scenario exploration is more valuable than provable correctness and safety, such as long-term capacity planning, demand forecasting, or policy evaluation. In such scenarios, AI surrogates offer an advantage by enabling faster-than-real-time simulation of a high number of scenarios.
In some cases, a hybrid approach is used where manually-built models act as a reference model and surrogates are deployed for fast approximation, with occasional recalibration to ensure validity as the system evolves.

\subsubsection*{Automated inference of digital twin components}

AI and machine learning can be used in the development phase of digital twins, too.
Developing simulators for digital twins is a pertinent example. Simulators are computer programs that encode an in-silico probabilistic mechanism and enact its changes over a sufficiently long period of time~\cite{ross2012simulation}. As such, they are key enablers of digital twins. However, simulator engineering is hindered by systems complexity~\cite{spiegel2005case}. Reusing and composing existing simulator components has been the traditional approach to building complex simulators. However, such techniques are hindered by vertical challenges stemming from inappropriate abstraction mechanisms, horizontal challenges stemming from different points of views, and increased search friction due to the abundance of information~\cite{page1999observations}.
To overcome the challenge of manual simulator construction, automated inference methods, e.g., based on reinforcement learning (RL) have been suggested.
RL is a machine learning mechanism that works by trial-and-error. The RL agent observes the workings of its environment in reaction to the agent's action. By that, the agent learns beneficial actions in specific situations. In simulator engineering, an RL agent can learn simulator input-output mappings and encode it in its policy, which subsequently can be used as a surrogate model. Integrated into a digital twin, an RL agent can actively experiment with its surroundings with reasonable boundaries of safety. For example, an RL agent in a digital twin of a smart grid might change the behavior of the grid as it optimizes it, and learn whenever an optimization action has turned out to be beneficial. RL can augment a digital twin and implement a continuous learning and calibration mechanism to keep simulators up-to-date~\cite{david2023automated}.

\subsubsection*{AI simulation}
AI simulation is the technique of applying AI and simulation technology to develop AI agents and the simulated environments in which they can be trained, tested and sometimes deployed~\cite{aisim-gartner}. 
A fundamental challenge in modern AI development is the scarcity of high-quality training data, particularly in domains where real-world data collection is expensive, dangerous, or impractical~\cite{zhou2017machine, farahani2023smart}. To overcome this issue, high-fidelity simulators can be used to generate synthetic training data for AI components. 

Digital twins are particularly well-suited for AI simulation due to their unique characteristics~\cite{liu2025ai}. First, the emergence of DTs elevated the quality, fidelity, faithfulness, and performance of simulators. These well-performing and high-fidelity simulators align well with the goals of AI simulation. 
Second, digital twins can support both passive observation and active experimentation with their physical counterparts, allowing for purposeful data collection that addresses specific AI training needs. The sophisticated simulation capabilities inherent in digital twins can generate diverse scenarios and edge cases that would be difficult or impossible to encounter in real-world operations.
The primary advantage of AI simulation lies in the ability to generate large volumes of high-quality, labeled training data under controlled conditions, while preserving the fidelity required for effective sim-to-real transfer. Digital twins support this process through either rapid, small-batch interactions or large-scale data generation. Moreover, digital twin-enabled AI simulation extends beyond simple data generation and labeling, enabling the creation of rich, interactive virtual training environments. Notably, AI simulation applications are predominantly built on genuine digital twins---those with bidirectional data and control flows. This underscores the importance of fully realized digital twin capabilities for effective AI training.

Digital twin-enabled AI simulation has proven to be particularly useful in complex domains such as wireless networks and robotics, where the combination of system complexity and sparse real-world data makes traditional AI training approaches challenging.
For smart grid applications, this approach can enable safe training of control algorithms and optimization strategies without risking disruption to critical power infrastructure.

\subsection{Digital Twins and Sustainability}

Sustainability is the ability to endure~\cite{lago2015framing} and maintain the function of a system over extended period of time~\cite{hilty2006relevance}. 
Driven by the increasingly more pronounced preference of individuals and society at large for human-centered, ecological, and economically viable systems, sustainability is becoming a key characteristic and driver of value in modern engineered systems, motivating companies to incorporate its various forms in their value propositions and operations, fostering circularity~\cite{david2024circular} and value retention across value chains~\cite{stucki2024data}. Sustainability of our technical and socio-technical systems has now become a strategic goal for governments as well, stimulating an array of international undertakings and innovation programs, e.g., the Industry 5.0 initiative of the European Commission.\footnote{\url{https://research and-innovation.ec.europa.eu/research-area/industrial-research-and-innovation/industry-50en}}
Digital twins have opened new frontiers for systems sustainability and excel at implementing various sustainability ambitions. Digital twins improve on traditional modeling capabilities by putting models into action~\cite{bork2024role}. The central tenet of this book is that digital twins enable more sustainable power generation as key components of smart grids. In addition, digital twins can control physical systems for various sustainability goals, rendering the system itself sustainable.

Digital twins contribute to sustainability in smart grids along environmental, economic, and social dimensions. 
From an \textit{environmental} perspective, digital twins allow grid operators to dynamically and automatically optimize the energy mix by simulating different dispatch strategies and enacting the optimal one. By that, digital twins help maximize the use of renewable energy sources and minimize the overall carbon intensity of the grid. For instance, by predicting solar and wind generation profiles across the grid, and aligning them with real-time demand, digital twins help ensure that clean energy is consumed when available, thus reducing dependency on fossil-based balancing mechanisms. 
From an \textit{economic} perspective, digital twins enable the predictive maintenance of critical assets, such as transformers, storage units, and transmission lines. Through the continuous monitoring of equipment health and failure forecasting, digital twins reduce unplanned downtime and associated costs. Optimized asset utilization extends the operational lifetime of expensive infrastructure elements, too.
Finally, from a \textit{social} perspective, digital twins contribute to the resilience and reliability of the grid, which reduces outage times and stabilizes energy access for communities. Digital twins enable the simulation of fault scenarios and stress conditions, which help operators prepare mitigation strategies in advance. In addition, digital twins can be used to evaluate and enact policies for equitable energy distribution, ensuring that vulnerable or remote populations benefit equally from sustainable energy.

Apart from implementing sustainability by digital twins, it is also important to implement digital twins in a sustainable way. In terms of a digital technology, such as digital twins, primary sustainability considerations include technical sustainability (e.g., evolution~\cite{david2023towards}), environmental sustainability (e.g., energy-efficiency~\cite{bellis2022challenges}), and social sustainability (e.g., situating the human in the loop~\cite{david2024infonomics}).
Digital twins are prime enablers of high-level legislative frameworks aiming at sustainable development. For example, twin transition~\cite{shajari2024twin} advocates identifying joint digital and sustainability transformation pathways that mutually amplify each other. Digital twins improve information flow across organizational silos and support decision-making about sustainability goals. Circular systems engineering~\cite{david2024circular} situates digital twins as a key enabler in the design, operation, and retirement of complex systems and also putting forward that at the same time, digital twins must become sustainable in order to avoid defeating their very purpose (i.e., to improve the sustainability of complex systems). In response to such emerging needs, the intensity of research both in digital twins for sustainability~\cite{heithoff2023digital,xu2022digital} and sustainability of digital twins~\cite{bellis2022challenges,david2023towards,fur2023sustainable} has increased recently.

\section{Concluding remarks: the future of digital twin-enabled smart grids}

This chapter provided an introduction to the foundations of digital twins and made the case for employing them in smart grids.
As engineered systems become more complex and autonomous, digital twin technology gains importance as the unified technological platform for design, testing, operation, and maintenance. Smart grids are prime examples of such complex systems, in which unique design and operation challenges arise from the combination of physical and software components. As high-fidelity in-silico replicas of physical components, digital twins provide safe and cost-efficient experimentation facilities in the design and verification phase of smart grids. In the operation phase of smart grids, digital twins enable automated load balancing of grids through real-time simulation and decision-making. These, and an array of similar benefits, position digital twins as crucial technological components in smart grids.

There are some key technological and methodological trends that are expected to shape of the digital twin landscape in smart grids.
First, the increasing adoption and deployment of edge computing infrastructure, along with the increasing computing power of edge nodes is expected to allow digital twins to be hosted closer to physical assets. This, in turn, will enable localized and distributed decision-making -- a critical capability for fast-changing grid conditions.
Second, trends in digital twins point towards aggregated and integrated digital twins~\cite{adesanya2024systems}, which is expected to give rise to full system-of-systems grid twin composed of individual digital twins of assets. System-of-system principles, such as subsystem independence and ad-hoc dynamic cooperation of constituent systems, align exceptionally well with the need to understand hard-to-model emergent behaviors of grids.
Finally, cognitive digital twins, i.e., digital twins that learn to self-optimize, are expected to become key enablers of the next generation of smart grids. With accessible AI solutions and affordable and mature hardware technology to support it, cognitive digital twins will contribute to more predictable, stable, and efficient grids.

\section*{Acknowledgment}
We acknowledge the support of the Natural Sciences and Engineering Research Council of Canada (NSERC), DGECR-2024-00293 (End-to-end Sustainable Systems Engineering).

\bibliographystyle{ACM-Reference-Format}
\bibliography{references}

\end{document}